# A new Contrast Based Image Fusion using Wavelet Packets

R. Balasubramanian and Gaurav Bhatnagar
Department of Mathematics,
Indian Institute of Technology Roorkee, Roorkee, India.
e-mail: balarfma@iitr.ernet.in, goravdma@iitr.ernet.in

*Abstract*—**Image Fusion, a technique which combines complimentary information from different images of the same scene so that the fused image is more suitable for segmentation, feature extraction, object recognition and Human Visual System. In this paper, a simple yet efficient algorithm is presented based on contrast using wavelet packet decomposition. First, all the source images are decomposed into low and high frequency sub-bands and then fusion of high frequency sub-bands is done by the means of Directive Contrast. Now, inverse wavelet packet transform is performed to reconstruct the fused image. The performance of the algorithm is carried out by the comparison made between proposed and existing algorithm.**

## I. INTRODUCTION

The information science research associated with the development of sensory system focuses mainly on how information about the world can be extracted from sensory data. In general, a single sensor is not sufficient to provide an accurate view of the real world. For the improvement in the capabilities of the intelligent machines and systems, concept of multiple sensors was presented. As a result, in the past few years multi-sensor fusion has become an important area of research and development. Hence, the single representation of different sources of sensory information is called multi-sensor fusion. Multi-sensor fusion can occur at the signal, image or feature. At most all advanced sensors of today, produce images. For example optical cameras, millimeter wave (MMW) cameras, infrared cameras, x-ray imagers, radar imagers etc. So the information, which we are getting from the advance sensors, is in the form of images. In image-based application fields, image fusion has emerged as a promising research area. Hence, image fusion is the process by which we combine two or more images into single image having important features from all. This fused image contains a more accurate description of the scene than any of the individual source images.

The simplest way for image fusion is pixel-by-pixel [1] gray level average of the source images. However this way leads to undesirable side effects such as reduced contrast. In the recent years, many image fusion methods have been proposed, such as statistical and numerical methods, hue-saturation- intensity (HSI) method, principal component analysis (PCA) method, image gradient pyramid[2,3] and multiresolution methods[4,5,6,8]. Statistical and numerical methods involve huge computation using floating point arithmetic. So these methods are time and memory consuming. The HSI method is based on the representation of the low spatial resolution images using HSI system and then substituting intensity component by a high resolution image. In PCA method, original images are transformed into uncorrelated images and then fused by choosing maximum value among all. PCA is frequently used for fusion because of its ability to compact the redundant data into fewer bands.

In the recent years, fusion methods based on image gradient pyramid and multiresolution analysis become very popular. The basic idea behind these methods is that source images are decomposed by applying pyramid or wavelet transform, then fusion operation is performed on the transformed images. These methods produce very good results in less computation time and less memory. Burt *et al.* [2] suggested a method in which the images are decomposed into gradient pyramid. Taking into account the variances in a 3×3 or 5×5 window, activity measure of each pixel is computed. Depending upon this measure, larger value is chosen. Li *et al.* [6] used similar method except the fact that for decomposition, discrete wavelet transformation(DWT) is used and consistency verification is also done along with area based activity measure and maximum selection. Pu *et al.* [8] suggested a contrast based image fusion method employed in the wavelet domain. After decomposition, directive contrast is computed for all decomposed images to fuse them.

In this paper, we present fusion scheme based on directive contrast using wavelet packet transform (WPT)[7,9] domain. In this scheme, we improve the method proposed by Pu *et al.*[8]. First we extend the concept of directive contrast for WPT domain (given in section 3). The benefit of WPT over DWT is that WPT allows better frequency localization of signals where we want as many small values as possible where as the standard wavelet transform may not produce the best result because it is limited to wavelet bases (the plural of basis). WPT increases by a power of two with each step. The comparison which is made by us shows that performance of our algorithm is better than Pu *et al.*[8].

The rest of the paper is organized as follows. The wavelet packet transform and directive constrast is explained in Section 2 and 3. In Section 4, proposed fusion algorithm is introduced. The experimental results are presented in section 5. Finally, the concluding remarks are given in Section 6.

## II. WAVELET PACKET TRANSFORM

The wavelet packet transform (WPT) generalizes the discrete wavelet transform and provides a more flexible tool for the time-scale analysis of data. All advantages of the wavelet transform are retained because the wavelet basis is in the repertoire of bases available with the wavelet packet transform. Given this, the WPT may eventually become a standard tool in signal and image processing.

Using a pair of low and high-pass filters to split a space corresponds to splitting the frequency content of a signal into roughly a low and a high-frequency components. In wavelet decomposition, we leave the high-frequency part alone and keep splitting the low-frequency part. Also in wavelet packet decomposition, we split the high-frequency part into a low and a high-frequency parts. So in general, wavelet packet decomposition divides the frequency space into various parts and allows better frequency localization of signals. Hence, WPT produces the complete binary tree(fig 1(b)).

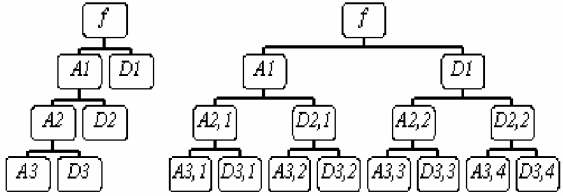

Fig. 1: Difference between DWT and DWPT a) DWT b) DWPT

### III. DIRECTIVE CONTRAST

According to *Human Visual System* (HVS) the local luminance contrast of images is defined as[3]:

$$C = \frac{L - L_B}{L_B} = \frac{L_H}{L_B} \qquad (1)$$

where $L$ and $L_B$ represent the local luminance and the luminance of the local background. Generally, $L_B$ is regarded as local low frequency and hence, $L-L_B=L_H$ is treated as local high frequency. On the above discussion a modified sequence of directive contrast for WPT is defined as:

Horizontal Contrast: $C_i^H = \dfrac{H_{i,\theta}}{A_l}$

Vertical Contrast: $C_i^V = \dfrac{V_{i,\theta}}{A_l}$ $\quad \begin{array}{l} 1 \le i \le l \\ \theta \in \{A,H,V,D\} \end{array}$

Diagonal Contrast: $C_i^D = \dfrac{D_{i,\theta}}{A_l}$

| $A_2$ | $H_{2,H}$ | $H_{1,A}$ | $H_{1,H}$ |
|---|---|---|---|
| $V_{2,V}$ | $D_{2,D}$ | $H_{1,V}$ | $H_{1,D}$ |
| $V_{1,A}$ | $V_{1,H}$ | $D_{1,A}$ | $D_{1,H}$ |
| $V_{1,V}$ | $V_{1,D}$ | $D_{1,V}$ | $D_{1,D}$ |

Fig. 2: 2 level DWPT of an image (Bolded block $A_2$ is the low frequency part for finding directive contrast)

### IV. PROPOSED FUSION ALGORITHM

In this section, we discuss some motivating factors in design of our approach to image fusion. We use DWPT and directive contrast for developing the algorithm. This scheme inherits advantages of both DWT and DWPT, i.e, better localization of low as well as high frequency. For our convinence, we take only two source images. Let they be *F1* and *F2*. For fusion, the basic condition is that the size of all source images are same. So without loss of generality let us consider, source images are of size $M \times N$. The fusion algorithm is given as follows:

*1)* First all the source images are *l*-level decomposed by the means of *discrete wavelet packet transform* (DWPT).

*2)* Find the sequence of directive contrast for each frequency. Let us denote $C_l^{\theta,1}$, $C_l^{\theta,2}$ are the directive contrast for first and second image respectively, where $\theta \in \{A,H,V,D\}$ and *l* is decomposition level.

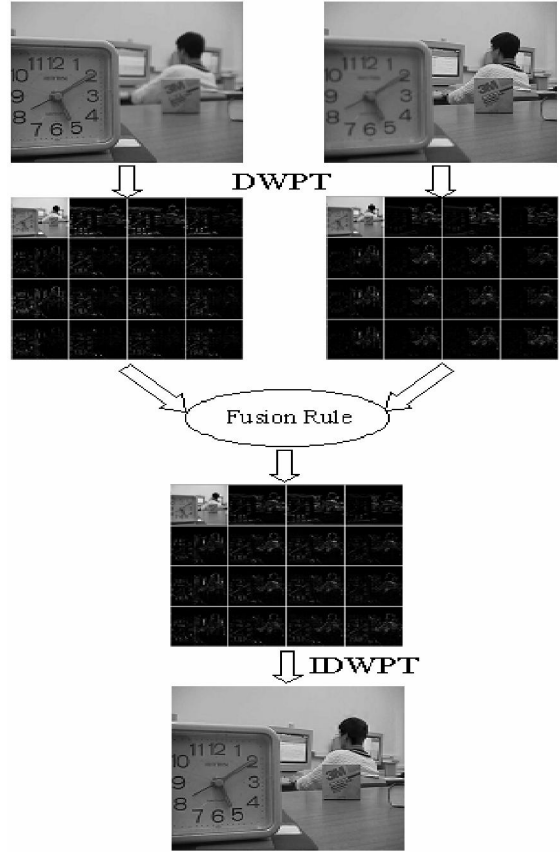

Fig. 3: Block Diagram of Proposed Algorithm

*3)* High frequency components are fused using directive contrast of corresponding pixels in all source images.

$$H_{i,\theta}^{new} = \begin{cases} H_{i,\theta}^1 & \text{if } \left|C_i^{H,1}\right| \ge \left|C_i^{H,2}\right| \\ H_{i,\theta}^2 & \text{otherwise} \end{cases}$$

$$V_{i,\theta}^{new} = \begin{cases} V_{i,\theta}^1 & \text{if } \left|C_i^{V,1}\right| \ge \left|C_i^{V,2}\right| \\ V_{i,\theta}^2 & \text{otherwise} \end{cases}$$

$$D_{i,\theta}^{new} = \begin{cases} D_{i,\theta}^1 & \text{if } \left|C_i^{D,1}\right| \ge \left|C_i^{D,2}\right| \\ D_{i,\theta}^2 & \text{otherwise} \end{cases}$$

where $1 \leq i \leq l$, $\theta \in \{A,H,V,D\}$ and $H_{i,\theta}^{new}$, $V_{i,\theta}^{new}$, $D_{i,\theta}^{new}$ are the new horizontal, vertical and diagonal components of the fused image.

*4)* For the fusion of low frequency (approximate part) we use median instead of averaging.

$$A_l^{new} = median(A_l^1, A_l^2)$$

*5)* Perform *inverse discrete wavelet packet transform* (IDWPT) to construct fused image.

## V. RESULTS AND DISSCUSSION

In our image fusion approach, first we establish an evolution index system. This system includes mean, standard deviation, entropy, average gradient, peak signal to noise ratio (PSNR) and correlation coefficients.

### A. Evaluation indices of image fusion

Image evaluation indices are used to evaluate the quality of the fused image. Definition of these indices and their physical meanings are given as follows.

*1) Mean and Standard Deviation:* In statistical theory, mean and standard deviation are defined as follows:

$$\hat{\mu} = \frac{1}{N}\sum_{i=1}^{N} x_i$$

$$\hat{\sigma}^2 = \frac{1}{N-1}\sum_{i=1}^{N}(x_i - \hat{\mu})^2$$

Where $N$ is the total number of pixels in the image and $x_i$ is the value of the $i^{th}$ pixel.

*2) Entropy:* Entropy is the measure of information quantity contained in an image. If the value of entropy becomes higher after fusion then the information quality will increase. Mathematically, entropy is defines as:

$$E = -\sum_{i=1}^{N} p(x_i) \ln p(x_i)$$

where $p(x_i)$ is the probability of the occurrence of $x_i$.

*3) Average Gradient:* The average gradient is given by:

$$\overline{g} = \frac{1}{N}\sum \sqrt{\frac{\Delta I_x^2 + \Delta I_y^2}{2}}$$

where $\Delta I_x, \Delta I_y$ are the differences in $x$ and $y$ direction. The larger the average gradient, the sharper the image.

*4) Peak Signal to Noise Ratio:* The PSNR indicates the similarity between two images. The higher value of PSNR is the better the fused image is.

$$PSNR = 10 \lg \frac{255^2}{RMSE^2}$$

Where RMSE (root mean square error) is defined as

$$RMSE^2 = \frac{1}{MN}\sum\sum[F_1(i,j) - F_2(i,j)]^2$$

*5) Correlation Coefficient:* The correlation coefficient is a number lies between [0, 1] that measures the degree in which two variables are linearly related. Correlation coefficient is given by:

$$\rho = \sqrt{\frac{\sum_{i=1}^{M}\sum_{j=1}^{N}[F_{ideal}(i,j) - F_{fused}(i,j)]^2}{MN}}$$

where $F_{ideal}$ is the ideal image and $F_{fused}$ is the fused image. Lower values of $\rho$ indicate greater similarity between the images $F_{ideal}$ and $F_{fused}$.

### B. Experimental Results

We have demonstrated the performance of the proposed fusion algorithm using MATLAB by taking different experimental images. We took mandrill, peppers, hoed, book and house as experimental images. All the images are of size 512×512 except hoed and gun, which are of size 256×256. In mandrill images, we concentrate on upper and lower half parts. In pepper images, we concentrate on left and right half parts. In hoed images, we concentrate on middle and outer parts. Gun images are the very famous example of the Concealed Weapon Detection. Home and book images are the examples of multi focus images. Further these two images are Color images. For all these images, we have compared our results with the results of Pu *et al.*[8].

For the fusion of color images, first we transform the color (RGB) image into Hue-Saturation-Intensity (HSI) image and then apply our algorithm on all the three parts of source images. Once we get fused hue, intensity and saturation parts, then apply inverse HSI transform to construct the fused RGB image. The procedure for RGB image fusion is given in figure 4.

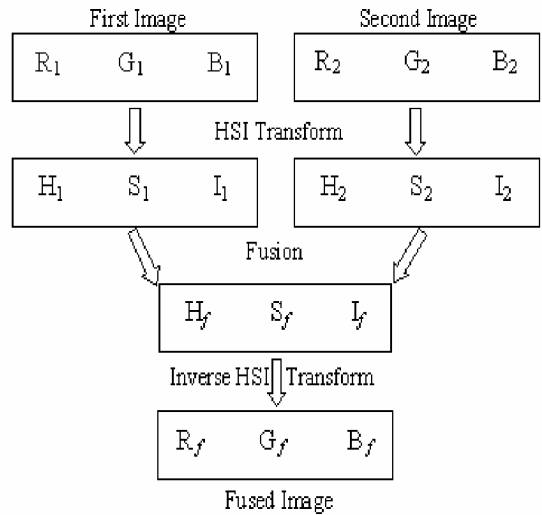

Fig. 4: Block Diagram for Fusion of Color Images

In figures 5,6,7,8,9 and 10, results of mandrill, pepper, hoed, gun, book and house images are given. The analysis which is done on the basis of evaluation indices is presented in table 1.

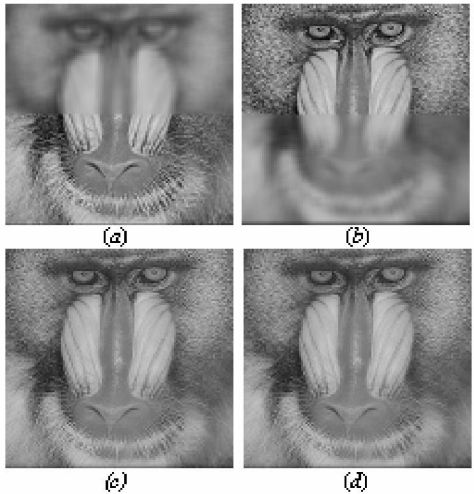

Fig. 5: Results for Mandrill image a) Upper side blurred image b) Lower side blurred image c) Result of DWT algorithm d) Result of DWPT algorithm

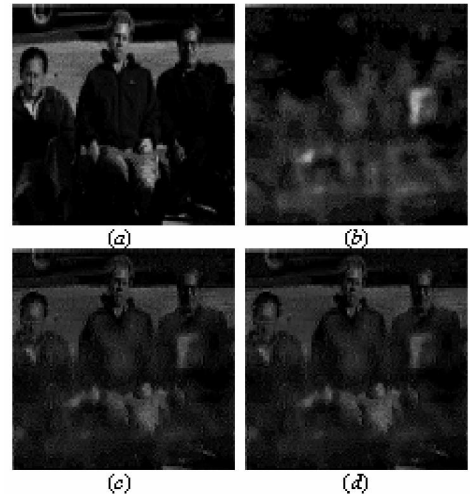

Fig. 8: Results for Concealed Weapon Detection a) Real image b) IR image c) Result of DWT algorithm d) Result of DWPT algorithm

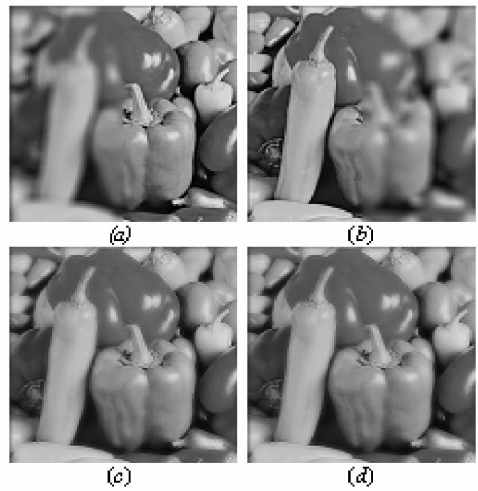

Fig. 6: Results for Pepper image a) Left side blurred image b) Right side blurred image c) Result of DWT algorithm d) Result for DWPT algorithm

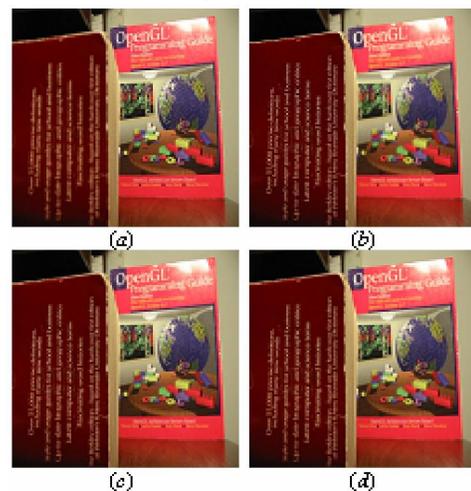

Fig. 9: Results for Book image a) Second book concentrated image b) First book concentrated image c) Result of DWT algorithm d) Result of DWPT algorithm

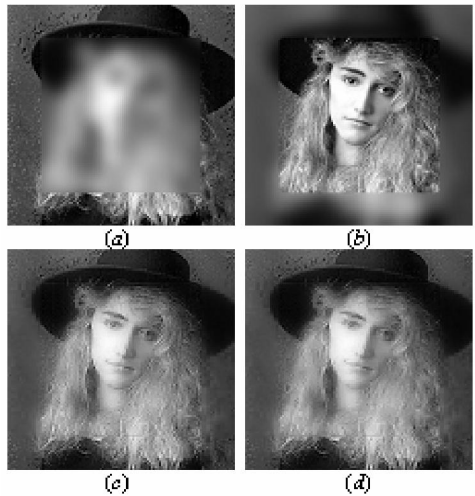

Fig. 7: Results for Hoed image a) Middle part blurred image b) Corner part blurred image c) Result of DWT algorithm d) Result of DWPT algorithm

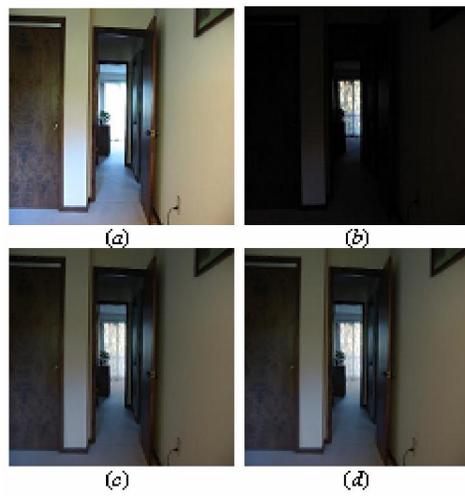

Fig. 10: Results for House image a) Brighter image b) Darker image c) Result of DWT algorithm d) Result of DWPT algorithm

It is clear from table 1 that our algorithm is performing better than Pu *et al.*[8]'s DWT method. Only for hoed image DWT method is performing better. For rest of the images DWPT method is performing better than Pu *et al.*[8]'s DWT method. We also compare mean, standard deviation, entropy and average gradient with original images and the values are very close to the original ones. Based on the experimental results obtained from this study, the wavelet packet based image fusion method is very efficient for fusing images. It shows better performance than wavelet transform based images.

## VI. CONCLUSSIONS

The fusion method described in this paper cover a large variety of practical applications. The presented fusion technique has been intended to help in understanding the current state of knowledge in this research area. In our proposed techniques we are using Wavelet Packet Transform instead of Discrete Wavelet Transform. The main benefit of WPT is that it allows better frequency localization of signals. Using this we produce as many small values as possible, according to our problem. The extended definition of directive contrast is mathematically more accurate than the common definition. It provides visually better fused images than existing algorithms. We demonstrate that the contrast-based wavelet packet fusion can do a better job than existing multiresolution fusion method[8], taking into account of mean, standard deviation, entropy, average gradient, PSNR and correlation coefficient.

## ACKNOWLEDGMENT

One of the authors, Gaurav Bhatnagar, gratefully acknowledges the financial support of the Council of Scientific and Industrial Research, New Delhi, India through his Junior Research Fellowship (JRF) scheme (CSIR Award no.: 09/143(0559)/2006-EMR-I) for his research work. The author, R. Balasubramanian, gratefully acknowledges the financial support of Sponsored Research and Industrial Consultancy (SRIC), Indian Institute of Technology Roorkee, India under the grant numbers FIG-100421-MAT and MHR03-05-802. He also acknowledges the financial support of Department of Science and Technology(DST), India under his fast track project for young scientist (DST-302-MTD) to carry out this research work.

TABLE 1.
EVALUATION INDICES FOR BOTH METHODS

| Image | | Mandrill | Pepper | Hoed | Gun | Book | House |
|---|---|---|---|---|---|---|---|
| Mean | Original | 129.6145 | 120.2164 | 95.9241 | — | 80.1558 | 63.8054 |
| | DWT | 129.3461 | 119.8029 | 95.9688 | 26.1908 | 84.0912 | 63.6090 |
| | DWPT | 129.3772 | 119.8287 | 96.0067 | 26.0583 | 84.1263 | 63.6306 |
| S.D. | Original | 5.8160 | 10.5246 | 19.3610 | — | 18.1151 | 16.8595 |
| | DWT | 5.4142 | 9.7441 | 19.7834 | 3.3262 | 17.4464 | 15.7182 |
| | DWPT | 5.3484 | 9.7856 | 19.5942 | 3.4156 | 17.5524 | 15.7667 |
| Entropy | Original | 5.1004 | 5.2635 | 5.3655 | — | 4.8840 | 4.8008 |
| | DWT | 5.0157 | 5.1952 | 5.3226 | 4.1391 | 4.9649 | 4.7749 |
| | DWPT | 4.9869 | 5.1941 | 5.3100 | 4.1593 | 4.9698 | 4.7833 |
| Gradient | Original | 28.4007 | 14.2918 | 20.0623 | — | 19.7677 | 8.3976 |
| | DWT | 25.4187 | 10.6006 | 14.7130 | 12.5426 | 17.2977 | 6.5125 |
| | DWPT | 22.8689 | 10.7870 | 12.3617 | 11.6551 | 17.0153 | 6.8209 |
| PSNR | DWT | 34.1722 | 35.5584 | 35.2760 | — | 36.8122 | 37.9599 |
| | DWPT | 34.2981 | 35.5592 | 33.9595 | — | 37.1661 | 38.1996 |
| C.C | DWT | 0.8722 | 0.8123 | 0.7992 | — | 0.7567 | 0.7347 |
| | DWPT | 0.8513 | 0.8120 | 0.8987 | — | 0.7124 | 0.7021 |

\* For the Gun image, proper original image is not available to compare our results. It is indicated by '—' in the table.